\documentclass[hidelinks,breaklinks,10pt,a4paper, twocolumn]{article}

\usepackage[utf8]{inputenc}%
\usepackage{amsmath}%
\usepackage{amsfonts}%
\usepackage{amssymb}%
\usepackage{makeidx} 
\usepackage{tikz}%
\usepackage{graphicx}%
\usepackage[labelfont=bf]{caption} 
\usepackage{accents}%
\usepackage{centernot}%
\usepackage{lmodern}
\usepackage{color}%
\usepackage[color,matrix,arrow]{xy}%
\usepackage{float}%
\usepackage[left=1.6cm,right=1.6cm,top=2cm,bottom=2.5cm]{geometry}%
\usepackage[style=phys,biblabel=brackets,sorting=none,backend=biber, doi=true,url=true, isbn=true,eprint=true]{biblatex}%
\DeclareFieldFormat{titlecase}{#1}
\usepackage{multicol}%

\title{\textbf{Efficient classical simulation of the Deutsch-Jozsa algorithm}}
\author{Niklas Johansson,$^*$ Jan-Åke Larsson$^{\dagger}$\\\small
  \textit{Institutionen för Systemteknik, Linköpings Universitet, 581 83
    Linköping, SWEDEN}} \date{}

\input{Qcircuit}

\begin{document}
\twocolumn[\maketitle
\begin{quote}
  In 1985, David Deutsch challenged the Church-Turing thesis by stating that
  his quantum model of computation \textit{``could, in principle, be built and
    would have many remarkable properties not reproducible by any Turing
    machine''}.  While this is thought to be true in general, there is usually
  no way of knowing that the corresponding classical algorithms are the best
  possible solutions. Here we provide an efficient classical simulation of the
  Deutsch-Jozsa algorithm, which was one of the first examples of quantum
  computational speed-up.  Our conclusion is that the Deutsch-Jozsa quantum
  algorithm owes its speed-up to resources that are not necessarily
  quantum-mechanical, and when compared with the classical simulation offers
  no speed-up at all.
\end{quote}
]


\noindent
The Deutsch-Jozsa algorithm \cite{Deutsch,Deutsch1992} was one of the first
indications of quantum computational speed-up \cite{Simon1994}. The algorithm
has, since then, been used extensively for illustrating experimental
realizations of a quantum computer and the quantum computational speed-up.
The size of this speed-up crucially depends on the complexity of the
corresponding most efficient classical algorithm. Here we present an algorithm
 within an
extension of Spekkens' ``toy theory'' (\cite{Spekkens}, here: toy model),
that  deterministically solves the Deutsch-Jozsa
problem with only one oracle query. Relative to the oracle, the algorithm is efficiently
simulatable on a classical Turing-machine, making it clear that there is
no quantum speed-up in this case. In the language of complexity classes, the
algorithm no longer gives any evidence of an oracle separation between EQP
(Exact or Error-free Quantum Polynomial time solvable problems
\cite{Bernstein1997}) and P (Polynomial time solvable problems).

Suppose that you are given a Boolean function $f(x):\{0,1\}^n \mapsto \{0,1\}$
with the promise that it is either constant or balanced. The function is
constant if it gives the same output (one or zero) for all possible inputs,
and it is balanced if it gives the output zero for half of the possible
inputs, and one for the other half. Your task is now to distinguish between
these two cases \cite{Deutsch1992,Cleve}.  Given such a function, a classical
Turing-machine can solve this problem by checking the output for $2^{n-1}+1$
values of the input; if all are the same, the function is constant, and
otherwise balanced. A stochastic algorithm with $k$ randomized function
queries gives a bounded error probability \cite{Deutsch1992} less than
$2^{1-k}$.

\begin{figure}[t]
\center
\includegraphics[scale=1.]{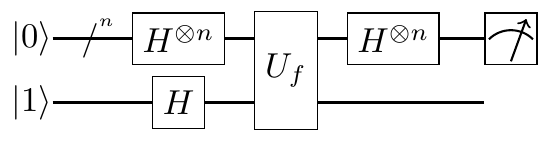}
\caption{\textbf{Circuit performing the Deutsch-Jozsa algorithm}. This circuit
  uses an $n$-qubit input-register prepared in the state $\ket{0}^{\otimes
    n}$, and a target prepared in $\ket{1}$. It proceeds to apply Hadamard
  transformations to each qubit. The function $f$ is embedded in an oracle
  $U_f$, and this is followed by another Hadamard transformation on each
  qubit. The measurement at the end will test positive for $\ket{0}^{\otimes
    n}$ if $f$ was constant, and negative if $f$ was balanced.}
\label{Circuit}\bigskip
\end{figure}

An oracle for a quantum computer, a quantum oracle, is a unitary
transformation that implements the function.  Given such an oracle,
a quantum computer can
solve the problem with a single query by using the Deutsch-Jozsa algorithm
\cite{Deutsch1992,Cleve}, in the ideal case with zero error probability.
Figure~\ref{Circuit} shows a quantum-circuit representation of the algorithm:
prepare an $n$-qubit input-register, a 1-qubit target, and put them through
Hadamard transformations to produce a full superposition over all
computational basis states. Perform the oracle transformation $U_f$,
corresponding to addition modulo 2 of the function value to the target qubit,
and finally apply Hadamard transformations to restore the input-register. This
procedure will leave the input-register unchanged only when the function is
constant, otherwise it will have changed \cite{Deutsch1992,Cleve}; a
measurement of the input-register will reveal this.

\begin{figure}[t]
\center
\includegraphics[scale=1]{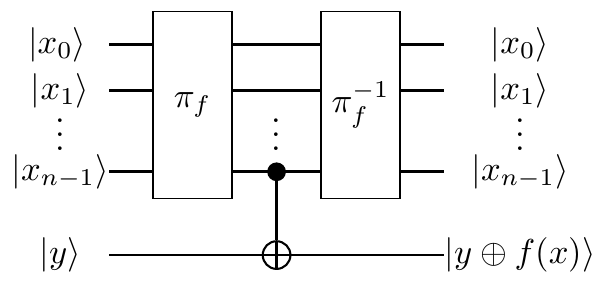}
\caption{\textbf{Oracle construction}. The $n$-qubit gates at the beginning
  and end are unitary permutations of computational basis states
  ($\pi_f^{-1}=\pi_f^{\dagger}$). At the center is a CNOT gate from the most
  significant qubit $\ket{x_n}$ to the target $\ket{y}$.}
\label{oracle}
\end{figure}

\begin{figure*}[t]
\center
\includegraphics[scale=.9]{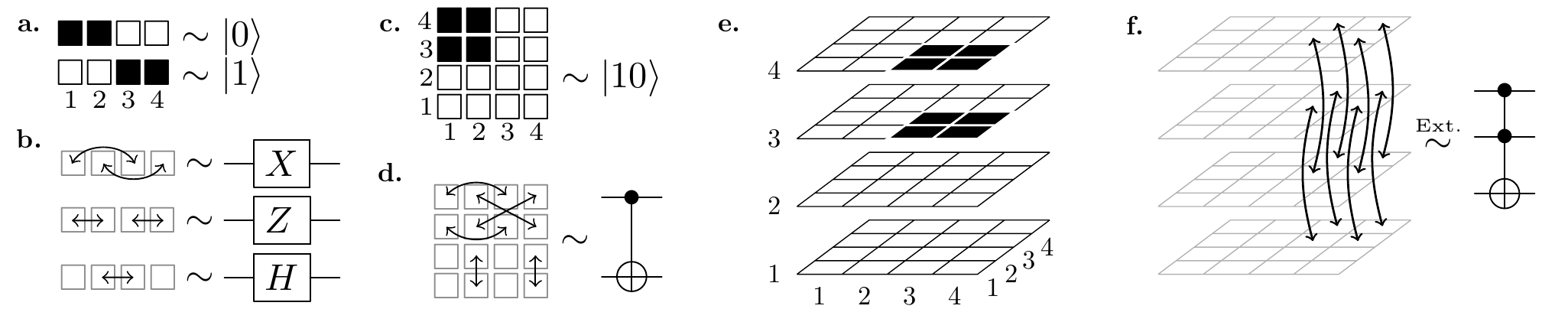}
\caption{\textbf{States and transformations.}  \textbf{a.}~Epistemic states of
  an elementary system encoding logical 1 and 0. \textbf{b.}~Permutations
  corresponding to Pauli-X, Pauli-Z, and Hadamard
  transformations. \textbf{c.}~Composite system of two elementary systems in
  an epistemic state corresponding to $\ket{01}$. \textbf{d.}~Permutation
  corresponding to a CNOT gate with the vertical system as control and the
  horizontal system as target. \textbf{e.}~A tripartite system in an epistemic
  state corresponding to $\ket{101}$, with the elementary system along the
  width and height as the most and least significant
  systems. \textbf{f.}~Transformation corresponding to the permutation
  $\ket{110} \longleftrightarrow \ket{111}$, note that this is not a valid
  permutation in Spekkens' toy model.}
\label{Spekkens_sys}
\end{figure*}

For $n=1$ and $n=2$ qubits in the input-register, the Deutsch-Jozsa oracle
is known to have an implementation that does not rely on quantum resources
\cite{Collins}.  These two cases can be implemented \cite{Niklas} in the
framework of Spekkens' toy model \cite{Spekkens} that captures many, but not
all, properties of quantum mechanics.  The toy model encodes the state of each
elementary system (or toy bit) into a one-out-of-four \emph{ontic state}. The
toy model follows the \emph{knowledge balance principle}; stating that at
every point in time at most one bit of information per toy bit can be known or
predicted by an observer. Therefore, a measurement can only yield knowledge
about which \emph{epistemic state} the system is in, corresponding to a
uniform probability distribution over the relevant part of the ontic state
space. The epistemic state is closely related to the quantum state. For
example, the epistemic state associated with $\ket0$ is a uniform distribution
over the ontic states 1 or 2, written $1\vee2$.  Each even partition of the
state space corresponds to a basis of the quantum state space as indicated in
Figure \ref{Spekkens_sys}a. Unitary transformations are modeled by ontic state
permutations, as indicated in Figure~\ref{Spekkens_sys}b. Systems compose
under the Cartesian product (see Figure~\ref{Spekkens_sys}c).

This construction even captures some properties of entanglement, enabling
protocols like super-dense coding and quantum-like teleportation, but cannot
give all consequences of entanglement, most importantly, it cannot give a Bell
inequality violation. However, the toy model's epistemic states and transformations
are closely related to stabilizer states and Clifford group
operations \cite{Matthew}, including the CNOT gate as shown in Figure
\ref{Spekkens_sys}d.  Indeed, a non-trivial subset of quantum mechanics can be
efficiently simulated on a classical Turing machine via the Gottesman-Knill
theorem \cite{Gottesman1998}. Furthermore, classical Turing
machine simulation of Spekkens' toy
model (of at most a polynomial size in the number of subsequent operations) is
efficient \cite{Spekkens}; all operations are time efficient, and the four ontic states of an
elementary system is stored in two classical bits; an $n$-toy-bit system
therefore requires $2n$ bits of storage.

However, for $n\ge3$, the Deutsch-Jozsa oracle needs the Toffoli gate, and
since the Toffoli gate is not efficiently simulatable using the stabilizer
formalism \cite{Gottesman1998} nor present in the toy
model \cite{Spekkens}, it has so far been believed that the Deutsch-Jozsa
algorithm does not have an efficient equivalent in the toy model.  Here, we
need to point out that our task is not to create Toffoli gate equivalents in
the toy model, or even simulate the quantum-mechanical system as such. It
suffices to give a working efficient toy model equivalent of the Deutsch-Jozsa
oracle. We therefore choose not to represent Toffoli gates exactly, but
design the ontic state permutation so that it swaps the epistemic states
associated with $\ket{110}$ and $\ket{111}$, but does not permute the ontic
states within these computational-basis epistemic states, see Figure
\ref{Spekkens_sys}f.  This type of transformations are, in general, not valid
transformations in Spekkens' toy model \cite{Spekkens}, but as already
stated, our aim is not to simulate quantum mechanics, it is to devise an
equivalent of the Deutsch-Jozsa algorithm. The extended model is still
efficiently simulatable on a classical Turing machine.

By representing a general computational-state permutation
\begin{equation}
  \pi =\sum_x \ket{\pi(x)}\bra{x}
\end{equation}
in the same manner as the Toffoli we obtain an extended toy model oracle,
proving the existence of such an oracle in this extended model. The toy model
permutation reminiscent of Hadamard gates \cite{Matthew} can now be used,
and the Deutsch-Jozsa algorithm works as in the quantum case.
Figure \ref{oracle} shows an oracle construction for balanced functions,
where $\pi_f$ is an arbitrary permutation of the computational basis states
which can be constructed from CNOT and Toffoli gates \cite{Nielsen}.  At the
center of the oracle a CNOT gate is applied between the most significant qubit
in the input-register and the target. With $\pi_f$ as the identity permutation, the
oracle performs the balanced function $f'$ that is 1 for all inputs with the
most significant bit set.  Any other balanced function can now be generated by
choosing a different $\pi_f$, giving the function output
$f(x)=f'(\pi_f(x))$. For a function that is constant
zero, the CNOT gate is omitted, so that the target value is unchanged
independently of the input-register.  The alternative constant one function
can be created by replacing the CNOT with a Pauli-X gate acting on the target,
inverting the target independently of the input-register.

This device will give the expected outcomes for classical function queries,
i.e., invocations that reveal a single function value. The desired input
number $x$ should be inserted in the input-register as the quantum state
$\ket{x}$ in the computational basis, and the target should be in a
computational basis state (say $\ket{0}$). Applying the oracle and measuring
the target will reveal the output of $f$: if the target was flipped, then the
function value is one for that input $x$.

The following is a simple description of the resulting complete toy-model
oracle in terms of a single transformation of the ontic state space. For any
balanced function the oracle will have the following effect:
\begin{enumerate}
\item The input-register ontic state belongs to some computational basis
  epistemic state associated with $\ket x$. If $f(x)=1$, perform a Pauli-X
  transformation on the target.
\item If the target system is in ontic state 2 or 4, perform a
Pauli-Z on the most significant input-register toy bit.
\end{enumerate}
The constant-function oracles do not have the CNOT, and will keep the input-register ontic state stationary for all possible inputs, because the
permutation $\pi_f$ is immediately followed by its inverse. The target ontic
state may change, depending on the actual constant function.

In our extended toy model algorithm we now prepare the input-register in
the epistemic state $(1\vee2)^n$ that corresponds to $\ket{0}^n$, and the
target in $3\vee 4$ that corresponds to $\ket1$, so that the whole system's
epistemic state is
\begin{equation}
  \big( \underbrace{1\vee 2, 1\vee 2, \ldots, 1\vee 2}_{n\text{
      elementary systems}}, 3\vee 4 \big) \ .
\end{equation}
The Hadamard transformation gives us
\begin{equation}
  \big( \underbrace{1\vee 3, 1\vee 3, \ldots, 1\vee 3}_{n\text{
      elementary systems}}, 2\vee 4 \big).
\end{equation}
We see that the target system is guaranteed to be in an even ontic state.

Applying the oracle for constant functions, the input-register will stay
unchanged, since the CNOT is not present, alternatively replaced with a
Pauli-X gate on the target toy bit. In this case, a second Hadamard
transformation on each toy bit will return the input-register to the initial
epistemic state, while there may be a change in the ontic state of the target
toy bit.

Applying the oracle for balanced functions the CNOT in the oracle will induce
a Pauli-Z permutation of the most significant system in the input-register.
\begin{equation}
 \big( \underbrace{2\vee 4, 1\vee 3, \ldots, 1\vee 3}_{n\text{
      elementary systems}}, 2\vee 4 \big)
\end{equation}
A second Hadamard transformation on the input-register will then produce the
epistemic state
\begin{equation}
 \big( \underbrace{3\vee 4, 1\vee 2, \ldots, 1\vee 2}_{n\text{
      elementary systems}}, 2\vee 4 \big)
\end{equation}

Measurement of the input-register will now reveal whether the function was
constant (the epistemic state completely overlaps with the initial epistemic
state) or balanced (the epistemic state is disjoint with the initial epistemic
state).

This system deterministically solves the problem using only one oracle
query. It can be efficiently simulated by a classical Turing machine with an
increase in memory (comparing qubits/toy bits with classical bits) by only a
constant factor of two. The spatial and temporal complexity of this simulation
are therefore identical to the complexity of the quantum algorithm. We do
recognize that this solution differs from the quantum solution in the sense
that the input epistemic state is mapped to the same output epistemic state
for all balanced functions, which is not necessarily the case with the quantum
solution.  Work has been done on a gate representation \cite{Niklas} that
mimics this behaviour more closely, but this does not reach zero error
probability, essentially because the quantum Toffoli gate does not have an
implementation in the toy model.

In conclusion, we have devised a toy model equivalent of the Deutsch-Jozsa
quantum algorithm, which is efficiently simulatable on a classical Turing machine. In the quantum algorithm, you are given an oracle that
implements a function $f$, your task is to distinguish balanced from constant
functions. In the presented classical simulation of the algorithm, you are given an oracle that
implements the the function in the same way, but within the framework of the extended toy model, and your task is still to distinguish balanced from constant
functions.  By this method we obtain equal temporal and spatial complexity as for
the quantum algorithm.  It is possible that the same technique can be used for
other quantum oracle algorithms, but these have been conjectured to use
genuinely quantum properties such as the continuum of quantum states, or
contextuality \cite{Larsson2012,Howard2014}, which both are missing from the
toy model \cite{Spekkens}. Therefore it remains to be seen what efficiency can
be achieved for these other algorithms. However, it is clear that the
Deutsch-Jozsa algorithm does not need genuinely quantum properties to work, and that the Deutsch-Jozsa algorithm shows no speed-up at
all compared with this new toy model algorithm or its classical simulation.\medskip

\emph{Acknowledgments.---} The project has been supported by the Foundational
Questions Institute (\href{www.fqxi.org}{FQXi}) through the Silicon Valley
Community Foundation.

\emph{Supplementary Information.---} Available in the form of a classical
simulation of the toy model algorithm, in Python, supplied with the online
version of the paper.

\printbibliography[heading=subbibliography,title={%
  \hfill\rule{.4\linewidth}{.3pt}\hfill\bigskip\\%
  \normalsize\mdseries%
  \null\hspace{1.1em}$^*$\quad Electronic address: \href{mailto:niklas.johansson@liu.se}{niklas.johansson@liu.se}\\
  \null\hspace{1.1em}$^\dagger$\quad Electronic address: \href{mailto:jan-ake.larsson@liu.se}{jan-ake.larsson@liu.se}}]




\end{document}